\documentclass[aps,prl,twocolumn]{revtex4}
\usepackage{xspace}
\usepackage{graphicx}

\newcommand{\pdreact}{$^2$H$(p,pd)$\xspace}

\newcommand{\thetacm}{\theta_\mathrm{cm}}
\newcommand{\etal}{\emph{et al.}\xspace}
\newcommand{\tmprime}{TM$^\prime$\xspace}
\newcommand{\urbnine}{Urbana-IX\xspace}
\newcommand{\chipt}{$\chi$PT\xspace}

\newcommand{\kviadd}{Kernfysisch Versneller Instituut (KVI), Zernikelaan 25, Groningen, The Netherlands}
\newcommand{\bochumadd}{Institut f\"ur Theoretische Physik II, Bochum, Germany}
\newcommand{\kyushu}{Department of Physics, Faculty of Engineering, Kyushu
Institute of Technology, Kitakyushu 804-8550, Japan}
\newcommand{\arizona}{Department of Physics, University of Arizona, Tucson,
AZ 85721, U.S.A.}
\newcommand{\cracowadd}{Institute of Physics, Jagellonian University, Cracow, 
Poland}

\begin{document}
\DeclareGraphicsExtensions{.eps}

\title{Systematic investigation of the elastic proton-deuteron differential
  cross section at intermediate energies}
\author{K. Ermisch$^{1,*}$}
\author{H.R. Amir-Ahmadi$^1$}
\author{A.M. van den Berg$^1$}
\author{R. Castelijns$^1$}
\author{B. Davids$^1$}
\author{E. Epelbaum$^2$}
\author{E. Van Garderen$^1$}
\author{W. Gl\"ockle$^2$}
\author{J. Golak$^3$}
\author{M.N. Harakeh$^1$}
\author{\mbox{M. Hunyadi$^1$}}
\author{M.A. de Huu$^1$}
\author{N. Kalantar-Nayestanaki$^1$}
\author{H. Kamada$^4$}
\author{M. Ki\v{s}$^1$}
\author{M. Mahjour-Shafiei$^1$}
\author{A. Nogga$^5$}
\author{R. Skibi\'nski$^3$}
\author{H. Wita{\l}a$^3$}
\author{H.J. W\"ortche$^1$}
\affiliation{$^1$\kviadd}
\affiliation{$^2$\bochumadd}
\affiliation{$^3$\cracowadd}
\affiliation{$^4$\kyushu}
\affiliation{$^5$\arizona}
\affiliation{$^*$ Present address: Onsala Observatory, Chalmers University of 
Technology, Gothenburg, Sweden}

\date{\today}

\begin{abstract}
  To investigate the importance of three-nucleon forces (3NF) systematically
  over a broad range of intermediate energies, the differential cross
  sections of elastic proton-deuteron scattering have been measured at
  proton bombarding energies of 108, 120, 135, 150, 170 and 190 MeV at 
  center-of-mass angles between $30^\circ$ and $170^\circ$. Comparisons with 
  Faddeev calculations show unambiguously the shortcomings of calculations 
  employing only two-body forces and the necessity of including 3NF. They 
  also show the limitations of the latest few-nucleon calculations at 
  backward angles, especially at higher beam energies. Some of these 
  discrepancies could be partially due to relativistic effects. Data at
  lowest energy are also compared with a recent calculation based on \chipt.
\end{abstract}

\maketitle

During the last decade, the addition of three-nucleon forces (3NF) to
modern high-quality nucleon-nucleon (NN) potentials, such as Nijmegen-I,
Nijmegen-II, Reid93, CD-Bonn and Argonne-V18 (AV18)
\cite{modernnn,modernnn2,modernnn3} has been shown necessary for many
three-nucleon scattering observables, like the differential cross section and
the vector analyzing power A$_y$ of elastic proton-deuteron scattering
\cite{witala1,bieber,sakai,ermisch,cadman,sekiguchi}. This necessity has 
been recognized before from the fact that 3N- and 4N- nuclei are 
underbound by NN forces alone \cite{anogga} and also because low-lying 
spectra of light nuclei can not be described correctly without 3NF
\cite{av18:new1,av18:new2}.  The most popular 3NF are \urbnine
\cite{urbanaix,urbanaorig} and a modified Tucson-Melbourne force \tmprime
\cite{3d,coonnew} (which no longer violates chiral symmetry).  Various
combinations of modern NN potentials and these 3NF have been worked out, 
fitted to the $^3$H binding energy \cite{anogga}, and subsequently applied
to 3N scattering. The comparison to data revealed a mixed picture. In some
cases NN forces alone work very well \cite{3ncomp}; in others, when NN force
predictions fail, the inclusion of 3NF leads to a good or fairly good
description \cite{witala1,bieber,sakai,ermisch,cadman,sekiguchi}, and in still
other cases neither NN forces alone nor the inclusion of these 3NF is
sufficient to describe the data
\cite{bieber,sakai,ermisch,sekiguchi,witala1,kuroszoln1,kuroszoln2}.
Recently, results of a systematic study of the nucleon vector-analyzing power
of the reaction \pdreact at several beam energies covering a large range of
center-of-mass angles were published \cite{ermisch}.  Serious discrepancies
were observed at backward angles, showing that the spin structure of 3NF is
not yet under control. Even though relatively precise data for the analyzing 
powers with large center-of-mass angular coverage exist in the literature
\cite{bieber,ermisch,sekiguchi,cadman}, for the differential cross section of
the reaction \pdreact few data sets are available. The existing data
\cite{postma,kuroda,adelberger,gigo}, with the exception of the data from
RIKEN \cite{sakai,sekiguchi}, are limited by low precision or small 
angular range.  Therefore, in order to obtain a comprehensive picture
of elastic proton-deuteron scattering at intermediate energies, a systematic
measurement of the differential cross section of this reaction as a function 
of the center-of-mass scattering angle and the bombarding energy was performed
at KVI.

The present data are compared to state-of-the-art solutions of the Faddeev
equations based on the above-mentioned high-precision NN potentials alone and 
in combination with the 3NF \tmprime and \urbnine. In addition, we shall 
include very recent results using NN and 3NF derived in chiral perturbation 
theory (\chipt). This new approach to nuclear forces is an extension of 
ongoing investigations in effective field theory applied to the nucleon 
itself, the $\pi$N and the $\pi$-$\pi$ systems. It is a
systematic perturbative approach based on a smallness parameter, the
ratio of generic external momenta to a certain mass scale of the order
of the $\rho$ mass. In the case of few-nucleon systems (and in the
Weinberg scheme \cite{3d}) the nuclear forces are expanded in relation to that
smallness parameter (and pion-mass insertions). These forces are then
inserted into the Schr\"odinger equation or equivalent formulations
like those of Faddeev-Yakubovsky. Recently, nuclear forces up
to next-to-next-to leading order (NNLO) have been worked out
\cite{eep02,eep02a}. At this order, NN forces consist of one- and
two-pion exchanges, which are parameter-free, and a string of contact
forces of low chiral dimensions. The parameters of the latter (so-called 
low-energy constants) have been fixed by NN scattering data \cite{eep01}. 
With these parameters, the forces can describe the data up to about 
200 MeV quite well. At this order 3NF begin to arise \cite{ukol94} and
consist of three different types of topologies: a) a parameter-free
two-pion exchange, b) a one-pion exchange between a nucleon and a
contact force between the other two, and c) a pure three-nucleon contact
force.  As has been shown in \cite{eep02a}, there are only two
parameters entering the 3NF of the types b) and c). They can be fixed
by adjusting to the $^3$H binding energy and to the doublet neutron-deuteron 
scattering length \cite{eep02a}.  This new dynamical picture has already been 
successfully applied in 3N- and 4N- systems, especially at lower energies for 
3N scattering up to 65 MeV \cite{eep01,eep02a}. It is now of great interest 
to see whether this approach will work even at 108 MeV, the lowest energy 
studied in this paper, or if higher orders in the chiral expansion or 
$1/M_N$ corrections are needed. Other approaches to calculate effective 
three-body forces exist in the literature \cite{Nemoto,canton}. At very low 
energies, techniques to solve the three-nucleon problem differently have 
been developed in recent years \cite{kievsky1,kievsky2}.

The experiment was performed at KVI using the Big-Bite Spectrometer
BBS \cite{bbs} in conjunction with the Euro-Supernova focal-plane
detection system ESN \cite{esn}. Protons were obtained from either
the KVI polarized ion-source \cite{polis} or the CUSP source and
accelerated in the superconducting cyclotron AGOR \cite{agor}.
Measurements were made at bombarding energies of 108, 120, 135, 150,
170 and 190 MeV. Since the lower acceleration limit of AGOR for
polarized protons is at 120 MeV, protons with a kinetic energy of 108
MeV were obtained by passing 120 MeV protons through an energy degrader.

When polarized ions were used, the polarization was continuously
measured using the KVI in-beam polarimeter (IBP) \cite{ibp} in the high-energy 
beam-line. For these measurements, analyzing powers were obtained in addition 
to cross sections. These analyzing powers were in total agreement with the 
published data \cite{ermisch} at the corresponding incident-beam energies.

As targets, mixed solid CD$_2$-CH$_2$ targets with several thicknesses and 
a well-known CD$_2$/CH$_2$ ratio were used. The target thickness was 
determined from measurements of the differential cross section of elastic
proton-proton scattering at several scattering angles. The measured
differential cross sections of this reaction were compared with the
results of NN calculations using Nijmegen-I, Nijmegen-II and
Reid93 potentials, resulting in a normalization factor for the
proton-deuteron differential cross sections. The analyzing power of the 
same reaction served as a cross-check of the polarization of the beam 
determined with IBP.

The BBS was positioned at laboratory scattering angles between
$17^\circ$ and $50^\circ$ in steps of $3^\circ$. At each scattering
angle, deuterons and protons emerging from the reaction \pdreact were
measured alternately. Deuterons were also measured at laboratory
scattering angles of $5^\circ$, $9^\circ$, $11^\circ$ and $14^\circ$.
In this way, center-of-mass angles between $30^\circ$ and $170^\circ$
were covered for all six incident-beam energies.

\begin{figure}[ht]
\begin{center}
\includegraphics[width=\linewidth]{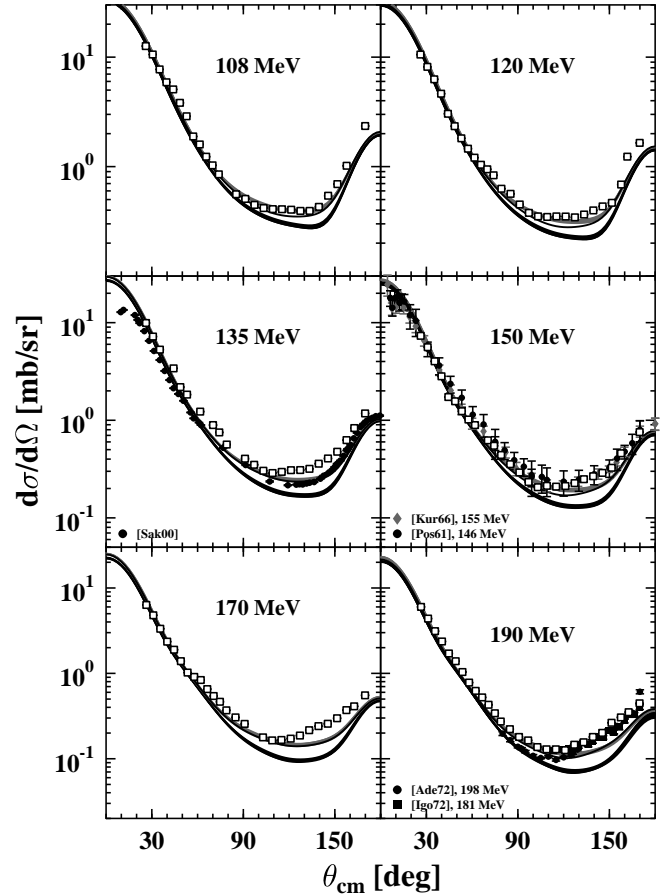}
\caption{\label{dsfig} Differential cross sections as a function
  of $\thetacm$. The data measured in this work are plotted as 
  open squares. The curves shown are calculations based solely on NN
  potentials (black band), calculations from Argonne-V18$+$\urbnine
  (solid line) and NN$+$TM$^\prime$ (gray band). At 135 MeV, data from 
  \cite{sakai} (circles), at 150 MeV data from \cite{postma} (circles) 
  and \cite{kuroda} (diamonds) and at 190 MeV, data from \cite{adelberger} 
  (circles) and \cite{gigo} (solid squares) are shown as well.}
\end{center}
\end{figure}

Measured differential cross sections are shown in figure \ref{dsfig}. 
The statistical uncertainty, which is in general of the order of $2\%$, is 
smaller than the size of the symbols. The total systematic uncertainty, which 
is the quadratic sum of the uncertainty in the normalization factor, the 
uncertainty in the polarization where polarized protons were used, and 
the uncertainty in the detection efficiency, is in general $\lesssim 7\%$, 
and should be considered as a point-to-point systematic uncertainty.

In all figures, the results of the Faddeev calculations using only 
two-nucleon interactions are shown as a black band. The width of 
this band represents the theoretical uncertainties in the calculations. 
Results from NN$+$\tmprime calculations are shown as a gray band. 
Further results from Argonne-V18$+$\urbnine are shown as solid lines. 
At 135 MeV, data from Sakai \etal \cite{sakai} are also shown. 
At 150 MeV, data from Postma \etal at 146 MeV \cite{postma} and
Kuroda \etal at 155 MeV \cite{kuroda} have been included in the
figure. At 190 MeV, data from Adelberger and Brown \cite{adelberger}
at 198 MeV and Igo \etal \cite{gigo} at 181 MeV are shown.
As can be seen, our data agree with most other existing data sets,
taking into account the experimental uncertainties and the difference
in incident-beam energy. At 135 MeV, our data lie systematically 
above the data from reference \cite{sakai}. 

\begin{figure}[ht]
\begin{center}
\includegraphics[width=\linewidth]{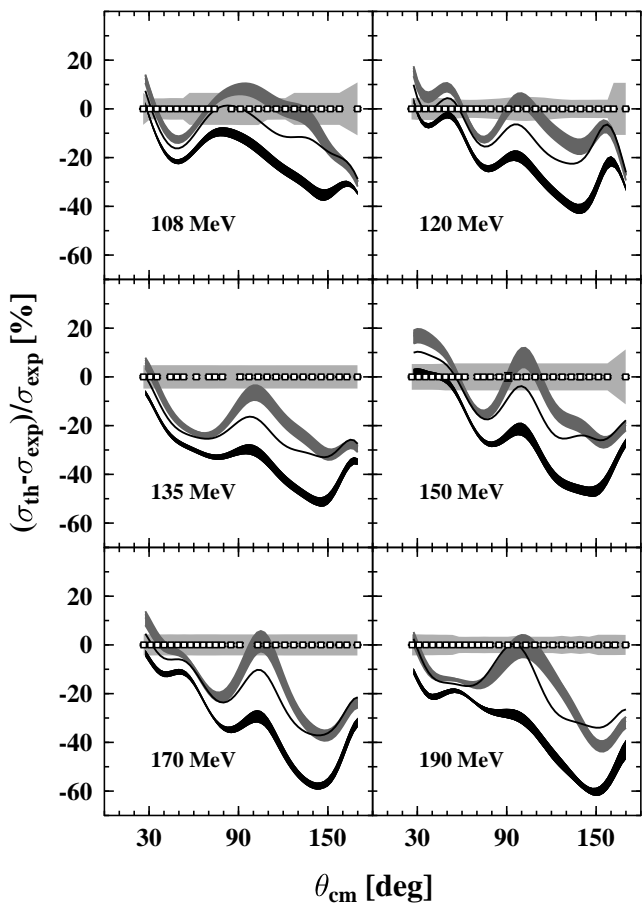}
\caption{\label{difffig} Relative deviations of modern NN and NN$+$3N 
  calculations from our data. The data points (open squares) lie at 
  zero with the statistical uncertainties denoted at each data point. 
  The total systematic uncertainty is shown as a light-gray band around zero. 
  The meaning of the curves is the same as in figure \ref{dsfig}.}
\end{center}
\end{figure}

To make a better comparison between our data and the theoretical
calculations shown in Fig.~\ref{dsfig}, the relative difference between 
them is shown in figure \ref{difffig}. In this figure, our data lie at 
zero and the calculations are shown by their relative deviations from our 
data. To avoid local fluctuations, these deviations are calculated by using
a polynomial fit through the data points.

In general, predictions based solely on NN interactions deviate from
our data not more than $\approx 60\%$ over a range of three orders of
magnitude. Inclusion of three-nucleon forces in the calculations
leads to improved agreement at all incident-beam energies. 
With the high-precision data obtained in this work, which
covers a broad center-of-mass angular region for six different
incident-beam energies, one can now study systematically the finer
details of the calculations.

At angles $\thetacm\approx 30^\circ$, almost all calculations overestimate the 
data. This is probably due to Coulomb effects not accounted for in the 
calculations. For $30^\circ\lesssim\thetacm\lesssim 60^\circ$, the predictions 
from different calculations with and without 3NF show some small
disagreement among themselves. Furthermore, the disagreement
between the theoretical predictions and our data is slightly outside
the systematic uncertainties. At angles $\thetacm\gtrsim 60^\circ$,
the predictions start to deviate from each other and from the data.
In a large part of the angular range, especially at the higher bombarding 
energies, calculations using solely two-nucleon potentials fail to
describe our data.  This deviation is largest around
$130^\circ\lesssim\thetacm\lesssim 150^\circ$.  This angular range is
part of the region of the minimum in the differential cross section
and the place where three-nucleon force effects are expected to
be largest \cite{witala1}.  The inclusion of three-nucleon forces
in the calculations remedies these discrepancies, especially at the
lower energies. However, starting at 135 MeV, deviations at backward
angles around $\thetacm\approx 140^\circ$ can be observed. These
deviations increase with increasing the bombarding energy. Furthermore, 
a local minimum in the difference plot in figure \ref{difffig} can 
be observed around $\thetacm\approx 70^\circ$. This minimum is due to a 
shoulder that begins at $\thetacm\approx 60^\circ$ and which is reproduced 
by the calculations but is more pronounced in our data. A similar pattern 
of disagreement exists in the vector-analyzing power \cite{ermisch}.  

\begin{figure}[ht]
\begin{center}
\includegraphics[width=\linewidth]{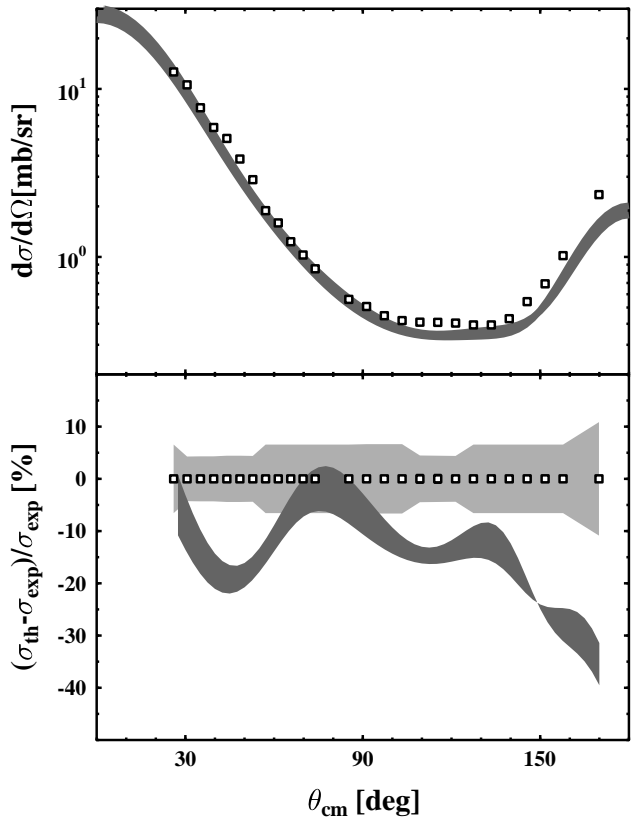}
\caption{\label{chipt} Absolute values and relative deviations of the 
calculated differential cross section in the framework of chiral perturbation
theory (dark-gray band) in comparison with the data at 108 MeV incident-beam 
energy. Also shown are the data points from this work and, in the lower panel, 
the experimental systematic uncertainty (light-gray band).}
\end{center}
\end{figure}

For the sake of clarity, we compare in Fig.~\ref{chipt} our results for the 
lowest energy, 108 MeV, to predictions of chiral perturbation theory. 
The theory is shown as a band, which reflects the dependence on a momentum 
cut-off parameter underlying the specific effective field theory formulation 
used. The agreement with the data at this order of the theory is comparable 
to the results based on the conventional forces. Higher-order corrections 
are expected to improve the description of the data.

The largest deviations of the theoretical predictions from our data
occur at high incident-beam energies and large backward angles, i.e.,
a kinematic regime where the momentum transfer is largest. A plausible
explanation for this is the higher-order effects, such as $\pi$-$\rho$ 
or $\rho$-$\rho$ exchange, which have not been included into the calculations. 
Part of the disagreement may also be due to relativistic corrections
\cite{kamada} which have not been properly taken into account in the 
calculations.

In conclusion, the high-precision data presented in this work, covering 
a large center-of-mass angular region at several bombarding energies, make 
possible systematic study of the influence and the deficiencies of 
existing three-nucleon forces. Calculations from NN$+$3N models show
deficiencies at backward angles at the higher bombarding energies. This may 
be an indication that the exchange of heavier mesons is missing in the 
calculations. It may also be an indication that further relativistic 
corrections must be included. The calculations based on chiral perturbation 
theory for an incident-beam energy of 108 MeV are promising and should be 
extended by incorporating N3LO and $1/m_N$ corrections.

\acknowledgements The authors wish to thank H. Meijer and his group at
the University of Groningen for the careful determination of the relative 
abundance of hydrogen and deuterium in the targets. They also had valuable 
discussions with R.G.E.~Timmermans. 
This work was performed as part of the research
program of the ``Stichting voor Fundamenteel Onderzoek der Materie''
(FOM) with financial support from the ``Nederlandse Organisatie voor
Wetenschappelijk Onderzoek'' (NWO) and was supported by the Deutsche
Forschungsgemeinschaft, the Polish Committee for Scientific Research 
under Grant No. 2P03B02818 and NSF Grant No. PHY0070858.  The numerical 
calculations of the Bochum-Cracow group have been performed on the Cray 
T90 and T3E of the NIC in J\"ulich, Germany.

\bibliography{collpapers}

\end{document}